\documentclass{winnower}

\begin{document}

\title{Should We Train Scientific Generalists?}

\author{Gopal P. Sarma, PhD}
\affil{School of Medicine, Emory University, Atlanta, GA, USA}

\date{}

\maketitle

\begin{abstract}
I examine the topic of training scientific generalists.  To focus the discussion, I propose the creation of a new graduate program, analogous in
structure to existing MD/PhD programs, aimed at training a critical mass of scientific researchers with substantial intellectual breadth.  In addition to completing the normal
requirements for a PhD, students would undergo an intense, several
year training period designed to expose them to the core vocabulary of
multiple subjects at the graduate level.  After providing some historical and philosophical context for this proposal, I outline how such a program could be implemented with little institutional overhead by existing research universities.  Finally, I discuss alternative possibilities for training generalists by taking advantage of contemporary developments in online learning and open science.
\end{abstract}

%-------------------------------------------------%
\section{Introduction}
%-------------------------------------------------%

%----------------------------------------------------------------------------------------
%	ESSAY BODY
%----------------------------------------------------------------------------------------

In the age of highly specialized science, the generalist is a long forgotten job description.  We have come to assume that the role played by those intellectual titans of earlier eras, such as Da Vinci, Aristotle, or Gauss, to name just a few, is an impossibility given the massive explosion of scientific knowledge in recent decades and centuries. \\

There is a factual reality to this sentiment that is uncontroversial.
Certainly, as a percentage of existing knowledge, one could not
conceivably attain the breadth of understanding that one might have in previous centuries.  However, it is
worth considering if a more modest goal could be achieved which would serve an important stabilizing role for modern science and
engineering.  That goal would be to train a critical mass of
scientific generalists, researchers, who in addition to the
specialized training of an ordinary graduate program, would also have broad exposure to multiple subjects at the graduate
level. \\

While the need for specialization might have been something of an inevitability, it is also worth
considering that there may be negative ramifications to this
kind of stratification of knowledge.  With so much to know, how can we
be confident that we are allocating our intellectual capital
efficiently?  How can we be confident in our collective understanding of global trends in science? \\

There is no doubt that in the coming years, data analytics of the
scientific corpus will play a significant role in contributing to the
creation of precisely such a global view of the scientific
enterprise.  The digitization of journals, the availability of
open API's for accessing scientific meta-data, and the integration of
reference management with social networking are all poised to
transform our understanding of the scientific process at a
high-level.  However, it seems naive to imagine that data mining techniques alone will allow
us to conceive of and test the most important hypotheses about the
global structure and dynamics of science without some
amount of guiding intuition.  To complement and maximally
take advantage of the availability of massive data sets about science,
as well as the computational tools to analyze those data sets, we need
a critical mass of scientific generalists whose training has been
designed to encourage hypothesis generation about the scientific
process itself. \\

Furthermore, another major trend in contemporary science is the move towards ambitious scientific agendas of substantially larger scope and project size \citep{nielsen2012reinventing}.  Whereas the pioneering theories of earlier eras were often crafted by solitary thinkers working in isolation, today's breakthroughs frequently come about from large international collaborations involving hundreds or thousands of people and research budgets in the billion dollar range.  In this context, the question of how to ideally train an individual scientist might be re-conceptualized as the question of how to train a scientific team member.  Scientific generalists could be pivotal members of such large collaborations and play critical organizational and leadership roles. \\

There are certainly scientific generalists today, although they are perhaps not thought of in this way.  I would broadly (and informally) categorize them into thee types:
\begin{itemize}
\item \textbf{The organic academic generalist}\\
This is someone who has led a traditional academic career on the tenure track, and whose research has naturally led to developing significant breadth in multiple topics.  Certainly many fields have researchers in this category.

\item \textbf{The academic-industrial wanderer}\\
This is someone who has left academia, or possibly had extended post-doctoral or research scientist appointments in subjects different from their PhD, and ultimately came back to academia, or led significant efforts at major industrial research laboratories.  For example, the growth of computational biology and theoretical neuroscience has been driven by many theoretical physicists who have gone on to do post-doctoral training in the biological sciences, or for example, physicists from the world of quantitative finance, who have returned to academia armed with a new set of skills quite different from their PhD training. 

\item \textbf{The autodidact}\\
The widespread availability of advanced scientific materials via the Internet has resulted in an organic trend towards the creation of generalists simply by lowering the barrier to accessing knowledge from a wide variety of fields, scientific or otherwise.  Certainly, there are many brilliant scientists in industry and elsewhere who do not have PhD's and it is not uncommon these days to encounter truly first class thinkers on a variety of topics who are largely self-taught.
\end{itemize}

The question that motivates this article is the following: should there be another category of generalist who has been trained from outset to play a different role in the modern scientific enterprise than researchers who set out to be specialists \citep{Bode03061949}?

\section{Training Scientific Generalists}  

As a means to stimulate discussion, but as an idea unto its own as well, I propose the following: the creation of a new graduate
program, roughly analogous in structure to an MD/PhD, where in
addition to the normal research requirements for completing a PhD, students
complete 5 or more qualifying examinations in subjects of their
choosing.  For adequately prepared students, I believe that after
completion of the requirements for their home department in their
first or second year, students would be able pass 4 additional
qualifying examinations over the course of 2-3 years, after which they
would resume their PhD research and complete their degree.\\

The choice of the qualifying examination as the focal point for this
program is that it encapsulates the basic vocabulary of a field, the
core knowledge required to conduct in depth research.
The aim of this program is emphatically \emph{not} to train researchers who have in
depth, specialized knowledge of 5 different subjects-- that would be
an unreasonable, if not outright impossible goal.  Rather, the aim is
to train students who understand the culture, the basic tools, and
broad perspectives of multiple subjects, so that they can contribute
to strengthening the very foundations of the scientific
establishment.  \\

Certainly, universities who undertake the process of creating such a program might choose to begin with a fewer number of qualifying examinations.  I chose this number because it would allow for individual students to engage with multiple, quite different subjects over the course of their graduate education, and because 6-8 month blocks per subject would create a program roughly on par with the length of an MD/PhD.  Part of the value in creating such a program would be the message and the vision it would send to younger students who are aspiring to life-long careers as scientists.  Just as undergraduates who aspire to careers as physician-scientists must adequately prepare themselves with appropriate exposure to both research and clinical work, aspiring scientific generalists would need to prepare themselves with advanced coursework of sufficient breadth to tackle the challenging initial years of this graduate program.  \\

For an ambitious program such as this one to maximally
benefit both the student and the scientific establishment at large,
there would need to be a strong culture to support those students who
choose to undergo such a rigorous and extended training.  In
particular, in order for the knowledge gained by these students to
develop into something much more rich and robust than a massive list
of facts and problem solving techniques from 5 different subjects,
they would need to be part of a mentoring program in which the process
of learning each of these different subjects was accompanied by
historical and philosophical discussion.  During each qualifying
examination block, students would ideally also attend regular seminars
in the department, and perhaps nominally be affiliated with a research
group and attend group meetings.  There would need to be a
culture among the students and faculty mentors that supported
reflection about problem solving strategies, about the structural
differences between the vocabulary and subject matter across different
fields.   Ultimately, these observations and insights, whether in raw
or more developed form would need to be communicated more broadly.\\  

One possibility might be to accompany the qualifying
examination process with a historical essay exploring some topic of
interest to the student in consultation with a faculty mentor.  For instance, a student whose PhD was in theoretical condensed matter physics and who passed examinations in physics, mathematics, chemistry, biology, and computer science might write an in depth essay on the emergence of quantitative methods in the study of natural selection.  A mathematics PhD student specializing in stochastic analysis and who passed qualifying examinations in mathematics, physics, statistics, computer science, and economics, might write about the contributions that mathematical finance pioneer Fisher Black made to macroeconomics.  \\

While this program may seem daunting, I would like to emphasize
that individuals who pursue MD/PhD degrees and ultimately become board
certified in a medical specialty need to pass a similar array of
hurdles-- in addition to PhD requirements for their research training
they have to pass multiple levels of board examinations to become
licensed physicians.  \\

It is also important to keep in mind that the training program described here is a graduate level training program, and consequently,
should be thought of as being the first step in a career-long
trajectory.  A person who completed this program is no more a mature scientific generalist than a person who completes an ordinary PhD program is a mature specialist.  In order for the subsequent phases of growth and development to take place, there would need to be a supporting infrastructure overseeing the post-doctoral period of the students' training.  Furthermore, it is certainly possible, and would be expected even, that a
subset of students who successfully complete this training would
simply choose to pursue tenure track jobs in their area of specialty.
Again, the MD/PhD is something of a guide-- certainly many dual-degree graduates
become purely clinical practitioners or pure researchers and do not
actively build careers bridging the two.  Students who pursue the more
traditional routes will not be at a disadvantage and one would hope that the unique and rigorous educational
experience they went through would inform the remainder of their
scientific careers both as researchers and as teachers.  But for those
that wish to mature into the novel role of scientific generalist that I am proposing, there would need to be special post-doctoral
programs providing generous several year funding that
would give them the freedom to develop their vision.  For the initial
batch of students, there would inevitably be some amount of trial and error while both
students and faculty develop an understanding of the strengths and
weaknesses of the program.  \\

While one can only speculate about the contributions graduates of this program
would ultimately come to make,  I close by suggesting a
few possibilities.  We might imagine tenured professorships for generalists who have
smaller research groups than they would otherwise have, but who are
active members of several different groups led by other faculty.  In addition to playing a
critical organizational role, these faculty members would bring
their considerable technical expertise and scientific breadth to each
group in the capacity of something along the lines of a scientific
consultant.  \\

Venture capital might be another place where scientific generalists
could have a significant impact, playing the
role of bridge builders between academia and industry, and perhaps
actively managing their own portfolios and overseeing scientific startup incubators.
\\

One of the most important roles generalists
could play would be to aid in the development of younger scientific
institutions, particularly in the developing world.  The specific aim
of this program is to train scientists who have significant exposure to the cultural elements of advanced science in multiple disciplines, whose training allowed them
to be both scientists as well as participatory anthropologists of the
scientific process.  For both younger universities in the developed
world, as well as new institutions in the developing world, scientific generalists could be
critical leaders and agenda setters, and perhaps, will be in a
position to identify important research trajectories, or important
cultural elements for executing those trajectories, that existing
institutions have overlooked.  \\

It would be incomplete and short-sighted to discuss novel educational initiatives without considering important contemporary developments in online education and open science.  Furthermore, given that another major contemporary theme in graduate education is the over production of PhD's relative to the availability of faculty positions, it would be understandable if a lengthy and extremely demanding variant of the PhD program is difficult to mobilize.  One possibility to balance these different factors would be to create an open system for crediting a student for having passed a qualifying examination.  Just as universities (and private companies) now offer certificates for coursework completed in a non-degree granting context, an open certification for anyone who is able to pass a qualifying examination would be a valuable credential that an individual could earn to demonstrate competency at the beginning graduate level.  For this certification to be available across all disciplines would be one step towards many different forms of educational innovation in the research world, including the training of scientific generalists.  \\

Before closing, let me re-examine the choice of the qualifying examination as the focal point for this particular proposal and consider alternatives.  Although the qualifying examination is an important rite of passage in graduate education, many will correctly point out that it is hardly something that contributes to depth of research maturity.  This is certainly a valid point, and in response, I would argue that the purpose of the program is not to train individuals who have achieved the maturity of the best specialists in multiple subjects, but rather individuals who can appreciate and communicate the knowledge of specialists, and who therefore would make strong collaborators, bridge builders, program managers, and journal editors etc.  The purpose of organizing a program for training generalists around the qualifying examination is in a large part analogous to why we have such examinations in the first place- they do contribute to some amount of intellectual and technical maturity and are an important experience to have early in one's education.  Furthermore, it is a simply stated idea that would require little institutional overhead, and would circumvent the inevitably controversial process of otherwise designing a curriculum.  \\

Also, it is worth mentioning that even within the same subject, there are many different types of qualifying examinations.  In the context of this article, Caltech's Computation and Neural Systems program (CNS) provides a possible template \citep{cns100}.  The model they employ is to give students a list of 100 questions used as preparatory material in the year leading up to an oral qualifying examination with 5 faculty members.  In a sense, the program is aimed at training ``generalists'' within the computational and neurobiological sciences.  It seems natural to ask if this model could be extended to incorporate other subjects as well.  That is, what if a list of 500 questions were to be assembled spanning multiple subjects and a set of  oral qualifying examinations were conducted by faculty spanning a number of different departments?  Or a few thousand questions from which a student selected some subset to prepare?   \\

It is not difficult to imagine alternatives, however.  One possibility would be a several year post-doctoral program where fellows rotated through several different laboratories and research groups in succession.  Or, in the spirit of the newly emerging trend of ``hacker schools'' and data-science boot camps, we could imagine creating analogously structured mini-courses designed by experts in the field targeted at advanced graduates whose training was in another field entirely.  Indeed, many academic research areas have highly topic specific summer schools and winter schools and one could imagine a several year post-doctoral program built around a handful of different sessions spread across multiple subjects.  Perhaps there should be a component both at the beginning of the PhD, like the hybrid qualifying examination system I outline above, as well as a post-doctoral component.  Ultimately, it is difficult to imagine that some trial and error would not be required in the design of such a program.  In addition, one thing is certain- successfully executing a program like this would require an organization to support students' growth for many years, and given the fundamentally experimental nature of such an effort, several years longer than we are accustomed to supporting graduate students.  Perhaps then, the ideal path forward would be to set into motion multiple efforts aimed at the common goal of training scientific generalists, so that over time, we can learn from our successes and mistakes.  To do so, of course, would require long-term institutional efforts to scientifically investigate the efficacy and impact of different training programs.  
\acknowledgments 
\subsection*{Acknowledgements}
I would like to thank Aaswath Raman, Doug Bemis, Venkatesh Narayanamurti, and Rob Spekkens for insightful discussions.    

\subsection*{\emph{Postscript}}
This article was written several years before \emph{The Winnower} was founded and was posted to several pre-print servers.  Because it is an editorial, as opposed to a research article, and so as to maintain consistency with other versions that have been online for some time now, I have chosen to freeze the initially posted version to \emph{The Winnower} and incorporate reviewer feedback into a subsequent series of articles on similar topics.  I hope members of the community will understand this decision, and I suspect that we are all just now beginning to appreciate the subtleties of post-publication peer review.  

\bibliographystyle{abbrvnat}
\bibliography{training_scientific_generalists}

\end{document}